\documentclass[12pt,a4paper]{iopart}

\usepackage[english]{babel}

\usepackage{iopams}
\usepackage{verbatim}

\usepackage{graphicx}

\usepackage{color}

\usepackage[colorlinks,citecolor=blue,linkcolor=black,urlcolor=blue]{hyperref}

\newcommand{\dd}{\mbox{d}}
\newcommand{\eq}[1]{(\ref{#1})}
\newcommand{\bun}{\hat{\mathbf{b}}}
\newcommand{\bv}{\mathbf{v}}
\newcommand{\bB}{\mathbf{B}}
\newcommand{\cE}{{\cal E}}

\newcommand{\rcoor}{{r}}

\begin{document}

%%\preprint{APS/123-QED}

\title[Stellarator impurity flux driven by electric fields tangent to magnetic surfaces]{Stellarator impurity flux driven by electric fields tangent to magnetic surfaces}

\author{Iv\'an Calvo$^{1}$}
\vspace{-0.2cm}
\eads{\mailto{ivan.calvo@ciemat.es}}
\vspace{-0.6cm}
\author{Felix I Parra$^{2,3}$}
\vspace{-0.2cm}
\eads{\mailto{felix.parradiaz@physics.ox.ac.uk}}
\vspace{-0.6cm}
\author{Jos\'e Luis Velasco$^{1}$}
\vspace{-0.2cm}
\eads{\mailto{joseluis.velasco@ciemat.es}}
\vspace{-0.6cm}
\author{J Arturo Alonso$^{1}$}
\vspace{-0.2cm}
\eads{\mailto{arturo.alonso@ciemat.es}}
\vspace{-0.6cm}
\author{J.~M. Garc\'{\i}a-Rega\~na$^{1}$}
\vspace{-0.2cm}
\eads{\mailto{jose.regana@ciemat.es@ciemat.es}}

\vspace{0.5cm}

\address{$^1$Laboratorio Nacional de Fusi\'on, CIEMAT, 28040 Madrid, Spain}
\address{$^2$Rudolf Peierls Centre for Theoretical Physics, University of Oxford, Oxford, OX1 3PU, 
UK}
\address{$^3$Culham Centre for Fusion Energy, Abingdon, OX14 3DB, UK}

%Uncomment for PACS numbers title message
\pacs{52.20.Dq, 52.25.Fi, 52.25.Xz, 52.55.Hc}
% Keywords required only for MST, PB, PMB, PM, JOA, JOB?
%\vspace{2pc}
%\noindent{\it Keywords}: Article preparation, IOP journals
% Uncomment for Submitted to journal title message
%\submitto{\PPCF}
% Comment out if separate title page not required

%%{\large
%%\begin{center}
%%\today
%%\end{center}
%%}

\begin{abstract}
The control of impurity accumulation is one of the main challenges for future stellarator fusion reactors. The standard argument to explain this accumulation relies on the, in principle, large inward pinch in the neoclassical impurity flux caused by the typically negative radial electric field in stellarators. This simplified interpretation was proven to be flawed by Helander \emph{et al.} [\emph{Phys. Rev. Lett.} {\bf 118}, 155002 (2017)], who showed that in a relevant regime (low-collisionality main ions and collisional impurities) the radial electric field does not drive impurity transport. In that reference, the effect of the component of the electric field that is tangent to the magnetic surface was not included. In this letter, an analytical calculation of the neoclassical radial impurity flux incorporating such effect is given, showing that it can be very strong for highly charged impurities and that, once it is taken into account, the dependence of the impurity flux on the radial electric field reappears. Realistic examples are provided in which the inclusion of the tangential electric field leads to impurity expulsion.
\end{abstract}

\maketitle

%%%%%%%%%%%%%%%%%%%%%%%%%%%%%%%%%%%%%%
\section{Introduction}
\label{sec:introduction}
%%%%%%%%%%%%%%%%%%%%%%%%%%%%%%%%%%%%%%

Stellarators~\cite{Helander2012} represent a promising alternative to tokamaks on the path towards a magnetic confinement fusion reactor. The main difference between a tokamak and a stellarator is that the former is axisymmetric, whereas the magnetic configuration of the latter is three-dimensional and (in general) has no symmetry direction. Due to the three-dimensional nature of the stellarator, the confining magnetic field can be generated by external coils only, whereas in a tokamak part of the magnetic field is produced by a large current in the plasma. This current drives plasma instabilities and makes it difficult to operate the device in steady-state, which is essential for a reactor. The intrinsic advantages of the stellarator regarding steady-state operation come at a cost: it is hard to design a three-dimensional magnetic configuration whose confinement quality is as good as that of the tokamak. Stellarator magnetic fields that are capable to perfectly confine particles in the absence of collisions, like tokamaks do, are called omnigeneous~\cite{Cary1997a, Cary1997b, Parra2015}. Omnigeneous stellarators have neoclassical transport levels (that is, transport due to inhomogeneities of the magnetic field and to collisions) comparable to those of tokamaks~\cite{Landreman2012}. However, even very small (and almost surely unavoidable) deviations from omnigeneity give large neoclassical transport, especially at low collisionality~\cite{Calvo2017}. The customary situation is that, in real stellarators, main ion transport is well described by neoclassical theory in the inner region of the plasma and turbulent transport is subdominant~\cite{Dinklage2013}. Highly charged ions are more collisional, but even for them neoclassical transport is expected to be very important. Understanding transport of impurity ions with high electric charge is particularly relevant because, for example, tungsten is the material planned for the divertor of ITER~\cite{Pitts2009} and is also the divertor material chosen in the conceptual design of some heliotron reactors~\cite{Sagara2017}.

In this letter we will study the neoclassical transport of impurities of mass $m_z$ and charge $Z_ze$ that collide with main ions of mass $m_i$ and charge $Z_ie$, where $e$ is the proton charge. We assume $Z_z \gg Z_i \sim 1$ and $\sqrt{m_z/m_i}\gg 1$. The neoclassical radial impurity flux across a magnetic surface (magnetic surfaces are often called ``flux surfaces") labeled by the radial coordinate $r$, $\Gamma_z$, can be written as
\begin{equation}\label{eq:Gamma_z_schematic}
\Gamma_z = -\eta (D_{11}^{zz} {\alpha}_{1z} + D_{11}^{zi} {\alpha}_{1i} + D_{12}^z {\alpha}_2),
\end{equation}
where $\eta(r)$ is related to the flux-surface average of the impurity density (see equation \eq{eq:def_Maxwellian} for its definition), $D_{11}^{zz}(r)>0$, $D_{11}^{zi}(r)$ and $D_{12}^z(r)$ are transport coefficients, and ${\alpha}_{1z} = \eta'/\eta + T'/T + Z_z e\varphi'_0 / T$,  ${\alpha}_{1i} = n_i'/n_i + T'/T + Z_i e\varphi'_0 / T$ and ${\alpha}_2 = T'/T$ are the thermodynamic forces. Here, $n_i(r)$ is the ion density, $T(r)$ is the temperature (the main ion and impurity temperatures are assumed to be equal) and primes stand for derivatives with respect to $r$. The electrostatic potential $\varphi(r,\theta,\zeta)$ is written as
\begin{equation}
\varphi(r,\theta,\zeta) = \varphi_0(r) + \varphi_1(r,\theta,\zeta),
\end{equation}
with $\theta$ and $\zeta$ poloidal and toroidal angles, respectively, and $|\varphi_1|\ll |\varphi_0|$. Hence, the radial electric field $E_r$ is approximately given by $E_r = -\varphi'_0$, whereas $\varphi_1$ determines the component of the electric field that is tangent to the magnetic surfaces.

In tokamaks, $D_{11}^{zi} = - (Z_z/Z_i) D_{11}^{zz}$ for arbitrary plasma profiles, and therefore $E_r$ does not appear on the right side of (\ref{eq:Gamma_z_schematic}). Moreover, $D_{12}^z$ is often positive and one can obtain $\Gamma_z > 0$ if $T'$ is sufficiently negative~\cite{Connor1973}; this is known as ``temperature screening". The situation is completely different in stellarators. Experimentally, stellarator impurity accumulation in the plasma core is consistently observed~\cite{Burhenn2009} (although there are remarkable, not completely understood, exceptions like the  ``impurity hole"~\cite{Ida2009, Sudo2016} in the Large Helical Device (LHD) and the ``High Density H mode" in Wendelstein 7-AS~\cite{McCormick2002}; see also \cite{Nakamura2017} for another exception in LHD involving plasmas with higher density than those typical of the impurity hole).  Impurity accumulation leads to fuel dilution and even to plasma termination by radiative collapse. This is why impurity transport is a matter of concern and an active research topic in the stellarator community. The standard theoretical argument to explain impurity accumulation, $\Gamma_z < 0$, in stellarators is as follows. Terms containing $E_r$ on the right side of (\ref{eq:Gamma_z_schematic}) do not cancel out and, since $E_r < 0$ in reactor-grade stellarator plasmas~\cite{Dinklage2013, Velasco2017}, the radial electric field gives a large inward pinch for $Z_z \gg 1$ (unless the value of $E_r$ is exceptionally close to zero~\cite{Velasco2017}).

Recently, it was proven in reference \cite{Helander2017} that the standard theoretical argument that we have just given to explain the phenomenon of impurity accumulation in stellarators is, sometimes, flawed. The calculation of \cite{Helander2017} shows that if $\varphi_1 \equiv 0$, the main ions have low collisionality and the impurities are collisional, then $D_{11}^{zi} = - (Z_z/Z_i) D_{11}^{zz}$ and, therefore, $E_r$ does not contribute to $\Gamma_z$. In principle, this is a positive result because it makes impurity screening in stellarators more likely than previously thought. The result of \cite{Helander2017} is correct and relevant as long as  $\varphi_1$ is negligible. The assumption that $Z_i e\varphi_1 / T$ is sufficiently small for $\varphi_1$ to be dropped in the main ion drift-kinetic equation is typically very good. However, in recent years it has become clear that, for $Z_z\gg 1$, the effect of $\varphi_1$ on the radial transport of impurities cannot be neglected~\cite{Garcia2017}, among other reasons, because the size of the radial $E\times B$ drift over the radial magnetic drift grows with $Z_z$. In this letter we assume the ordering
\begin{equation}
\frac{Z_z e \varphi_1}{T} \sim 1
\end{equation}
and show that, when the effect of $\varphi_1$ is incorporated, $E_r$ reappears on the right side of (\ref{eq:Gamma_z_schematic}). Actually, we show that the effect of $\varphi_1$ on $\Gamma_z$ can be large even if $Z_z e \varphi_1 /T$ is small. We prove all this by explicitly deriving in Section \ref{sec:derivation_Gammaz}, for trace impurities, an analytical expression for $\Gamma_z$ including the effect of $\varphi_1$, assuming that the ions have low collisionality, that the impurities are collisional (i.e. they are in the so-called Pfirsch-Schl\"uter regime) and that the electrons are adiabatic. This expression is provided in equation \eq{eq:Gamma_z_Fz1minus1_final}, which reduces to that in \cite{Helander2017} when $\varphi_1\equiv 0$. The result given in \eq{eq:Gamma_z_Fz1minus1_final} has been obtained in parallel and independently in reference \cite{Buller2018} by employing a fluid treatment for the impurities. The approach that we follow in this letter is fully kinetic. In Section \ref{sec:numerical_evaluation_Gammaz}, as an example, we provide a numerical evaluation of expression \eq{eq:Gamma_z_Fz1minus1_final} in a realistic LHD plasma.

%%%%%%%%%%%%%%%%%%%%%%%%%%%%%%%%%%%%%%%%%%%%%%%%%
\section{Derivation of the expression for the neoclassical radial impurity flux}
\label{sec:derivation_Gammaz}
%%%%%%%%%%%%%%%%%%%%%%%%%%%%%%%%%%%%%%%%%%%%%%%%%

In velocity space, we employ as coordinates the total energy per mass unit $\cE = v^2/2 + Z_z e\varphi/m_z$,
the magnetic moment $\mu = v_\perp^2 / (2 B)$, the sign of the velocity parallel to the magnetic field $\bB$, $\sigma = v_{||} / |v_{||}|$, with
$v_{||} = \sigma \sqrt{2\left(\cE -\mu B - 
{Z_z e} \varphi /m_z
\right)}\,\,$,
and the gyrophase $\phi$. Here, $v$ is the magnitude of the velocity $\bv$, $v_\perp$ is the magnitude of its component perpendicular to the magnetic field and $B=|\bB|$. The impurity distribution $F_z(r,\theta,\zeta,\cE,\mu)$, which is gyrophase-independent for the purposes of this calculation~\cite{Calvo13}, can be written as $F_z = F_{Mz} + F_{z1}$, where
\begin{equation}\label{eq:def_Maxwellian}
F_{Mz} =
{\eta}
\left(
 \frac{m_z}{2\pi T}
 \right)^{3/2}
 \exp\left(\frac{Z_z e\varphi_0}{T}\right)
 \exp\left(
-\frac{m_z \cE}{T}
\right)
\end{equation}
is a Maxwellian distribution constant on the magnetic surface, $\eta(r)$ is a function that depends only on the radial coordinate $r$ and $F_{z1}$ is a perturbation to $F_{Mz}$. Note that the density of the Maxwellian distribution, $n_z = \int F_{Mz}\dd^3 v$, is related to the functions $\eta$, $\varphi_1$ and $T$ by
\begin{equation}
n_z(r,\theta,\zeta) = {\eta}(r) \exp\left(-\frac{Z_z e \varphi_1(r,\theta,\zeta)}{T(r)}\right).
\end{equation}

In terms of the distribution function, $\Gamma_z$ reads
\begin{equation}\label{eq:def_Gamma_z}
\Gamma_z = \left\langle \int v_{d,r} F_{z1} \dd^3 v \right\rangle,
\end{equation}
where $\langle\cdot\rangle$ denotes flux-surface average, the radial drift $v_{d,r}$ is the sum of the radial components of the magnetic and $E\times B$ drifts,
\begin{equation}\label{eq:radial_drift}
\hspace{-0.15cm}
v_{d,r}
=
\frac{v_{||}}{\Omega_z}\nabla\cdot\left(v_{||}\bun\times\nabla r\right),
\end{equation}
and $\Omega_z = Z_z e B /m_z$ is the impurity gyrofrequency.

The function $F_{z1}$ is determined by the drift-kinetic equation
\begin{equation}\label{eq:DKE_impurities}
v_{||}\bun\cdot\nabla F_{z1} + v_{d,r} \partial_r F_{Mz} = C^{(\ell)}_{zi}[F_{z1};h_i].
\end{equation}
Here, $C^{(\ell)}_{zi}[F_{z1};h_i]$ is the linearized impurity-ion collision operator and $h_i$ is the non-adiabatic component of the deviation of the main ion distribution from a Maxwellian distribution. We have used the trace impurity approximation to neglect impurity-impurity collisions against impurity-ion collisions (impurity-electron collisions can be neglected against impurity-ion collisions due to the small mass of the electrons relative to the mass of the ions). To lowest order in a $\sqrt{m_i / m_z} \ll 1$ expansion~\cite{HelanderBook}, we have
\begin{eqnarray}\label{eq:C_zi_lin_expanded_in_mass_ratio}
C^{(\ell)}_{zi}[F_{z1};h_{i}] =
\nu_{zi}\left(
{\cal K} F_{z1}
+
\frac{m_z A v_{||}}{T} F_{Mz}\right),
\end{eqnarray}
where $
\nu_{zi} = 
{ Z_z^2 Z_{i}^2 e^4 n_i m_i^{1/2} \ln\Lambda}/{[3(2\pi)^{3/2}\varepsilon_0^2 m_z T^{3/2}]}
$ is the impurity-ion collision frequency, $\varepsilon_0$ is the vacuum permittivity, $\ln\Lambda$ is the Coulomb logarithm,
\begin{equation}
{\cal K}F_{z1} = \frac{T}{m_z}
\nabla_v\cdot
\left(
F_{Mz}
\nabla_v
\left(\frac{F_{z1}}{F_{Mz}}\right)
\right)
\end{equation}
and
\begin{eqnarray}\label{eq:def_vectorA}
A
=
\frac{3\sqrt{\pi\,}T^{3/2}}{\sqrt{2}n_i m_i^{3/2}}
\int
\frac{v_{||}}{v^3}
h_{i}(\bv)
\dd^3v
\end{eqnarray}
is the parallel velocity that the impurities would reach due to friction with the main ions in the absence of other forces. Hence, we would like to solve
\begin{eqnarray}\label{eq:dke_z_again}
&&\hspace{-0.7cm}
(v_{||}\bun\cdot\nabla
-
\nu_{zi}
{\cal K}) F_{z1}
 =
 \nonumber\\
&&
- v_{d,r} \partial_r F_{Mz}
+
\frac{\nu_{zi} m_z A v_{||}}{T} F_{Mz}
\end{eqnarray}
in the Pfirsch-Schl\"uter regime, defined by $\nu_{zi*} = \nu_{zi} R_0/ v_{tz} \gg 1$, where $\nu_{zi*}$ is the impurity-ion collisionality, $v_{tz} = \sqrt{2T/m_z}$ is the impurity thermal speed and $R_0$ is the stellarator major radius.

We expand $F_{z1}$ as $F_{z1} = F_{z1}^{(-1)} + F_{z1}^{(0)} + F_{z1}^{(1)} + \dots$,
with $F_{z1}^{(k)} \propto \nu_{zi*}^{-k}$. The terms in \eq{eq:dke_z_again} that scale with $\nu_{zi*}^{2}$ impose
\begin{eqnarray}\label{eq:dke_z_new_order_minus1}
{\cal K} F_{z1}^{(-1)}
= 0,
\end{eqnarray}
which implies
\begin{equation}\label{eq:Fz1minus1}
F_{z1}^{(-1)}(\rcoor,\theta,\zeta,{\cal E}) = \frac{N^{(-1)}(\rcoor,\theta,\zeta)}{n_z(\rcoor,\theta,\zeta)} F_{Mz}(r,{\cal E}).
\end{equation}
The contribution of $F_{z1}^{(-1)}$ to \eq{eq:def_Gamma_z} gives the neoclassical radial impurity flux in the Pfirsch-Schl\"uter regime. Using \eq{eq:Fz1minus1} in \eq{eq:def_Gamma_z}, taking the integral over velocities and defining a function $U_1(r,\theta,\zeta)$ as the solution of
\begin{equation}\label{eq:def_U1}
\bB\cdot\nabla U_1 = -\nabla\cdot
\left(
\frac{1}{\Phi_1}\frac{\bun\times\nabla r}{B}
\right),
\end{equation}
where we have introduced the short-hand notation $\Phi_1 \equiv \exp(Z_z e \varphi_1 / T)$, we get
\begin{equation}\label{eq:Gamma_z_Fz1minus1_4}
\Gamma_z = 
\frac{T{\eta}}{Z_z e}
\left\langle
U_1
\bB\cdot\nabla\left(
\frac{N^{(-1)}}{n_z}
\right)
\right\rangle.
\end{equation}
Denoting by $(r, \theta_{\rm max}, \zeta_{\rm max})$ the point where $B$ reaches its maximum value on the surface, we choose to fix the integration constant in \eq{eq:def_U1} such that $U_1(r, \theta_{\rm max}, \zeta_{\rm max}) = 0$. In order to determine $\bB\cdot\nabla(N^{(-1)}/n_z)$, we need higher order equations in the $1/\nu_{zi*}$ expansion.

Terms in \eq{eq:dke_z_again} that scale as $\nu_{zi*}$ give
\begin{equation}\label{eq:dke_z_new_order_zero_rewritten}
{\cal K} F_{z1}^{(0)}
=
\left[
\frac{1}{\nu_{zi}}
\bun\cdot\nabla
\left(
\frac{N^{(-1)}}{n_z}
\right)
-
\frac{m_z A}{T}
\right]
v_{||} F_{Mz},
\end{equation}
where \eq{eq:Fz1minus1} has been employed.
It is easy to check that 
\begin{equation}\label{eq:solutionFz1zero}
F_{z1}^{(0)}
=
-
\left[
\frac{1}{\nu_{zi}}
\bun\cdot\nabla
\left(
\frac{N^{(-1)}}{n_z}
\right)
-
\frac{m_z A}{T}
\right]
v_{||} F_{Mz}
\end{equation}
solves \eq{eq:dke_z_new_order_zero_rewritten}.

From terms in \eq{eq:dke_z_again} that scale as $\nu_{zi*}^0$, we find
\begin{eqnarray}\label{eq:dke_z_new_order_1}
&&{\cal K} F_{z1}^{(1)}=
\frac{1}{\nu_{zi}}
v_{||}\bun\cdot\nabla F_{z1}^{(0)}
\\
&&
+
\frac{1}{\nu_{zi}}
v_{d,r} \left[
{\alpha}_{1z} +
{\alpha}_2\left(\frac{m_z \cE}{T}-\frac{Z_ze\varphi_0}{T} - \frac{5}{2}\right)
\right] F_{Mz}
. \nonumber
\end{eqnarray}
We will not need to solve for $F_{z1}^{(1)}$ in \eq{eq:dke_z_new_order_1}, but only deal with its solvability condition, obtained by integrating over velocities. The solvability condition is
\begin{eqnarray}\label{eq:H2_solvability_condition}
&&\bB\cdot\nabla
\left(\int \frac{v_{||}}{B} F_{z1}^{(0)}
\dd^3 v\right) =
\\
&&
-
\int
v_{d,r} \left[
{\alpha}_{1z}
+
{\alpha}_2
\left(\frac{m_z \cE}{T}-\frac{Z_ze\varphi_0}{T} - \frac{5}{2}\right)
\right] F_{Mz}
\dd^3 v.
\nonumber
\end{eqnarray}

Using \eq{eq:solutionFz1zero} on the left side of \eq{eq:H2_solvability_condition} and working out explicitly the integral on the right side of \eq{eq:H2_solvability_condition}, the solvability condition becomes
\begin{eqnarray}\label{eq:solvability_condition}
&&\bB\cdot\nabla
\left\{
\frac{T}{m_z B}\frac{1}{\Phi_1}
\left[
\frac{1}{\nu_{zi}}
\bun\cdot\nabla
\left(
\frac{N^{(-1)}}{n_z}
\right)
-
\frac{m_z A}{T}
\right]
\right\}
 =
 \nonumber\\
&&
 \nabla\cdot
\left\{
\frac{T}{Z_z e}\frac{\bun\times\nabla r}{B}
\frac{1}{\Phi_1}
\left[
{\alpha}_{1z}
+
{\alpha}_2
\frac{Z_z e \varphi_1}{T}
\right]
\right\}
 .
\end{eqnarray}
This expression suggests the definition of another function $U_2(r,\theta,\zeta)$ via the equation
\begin{equation}\label{eq:def_U2}
\bB\cdot\nabla U_2 = -\nabla\cdot\left(
\frac{Z_ze\varphi_1}{T}
\frac{1}{\Phi_1}
\frac{\bun\times\nabla r}{B}\right)
\end{equation}
together with the condition $U_2(r, \theta_{\rm max}, \zeta_{\rm max}) = 0$. On an ergodic surface, equation \eq{eq:solvability_condition} implies that
\begin{eqnarray}\label{eq:solvability_condition_3}
&&\bB\cdot\nabla
\left(
\frac{N^{(-1)}}{n_z}
\right)
 =
\frac{\nu_{zi}m_z}{T}\left(B A -
\frac{B^2 \Phi_1}{\langle B^2 \Phi_1\rangle}
\langle B A \rangle\right)
\nonumber\\
&&\hspace{0.2cm}
+\frac{\nu_{zi} m_z}{Z_z e}\Bigg[
- B^2\Phi_1
\left(
{\alpha}_{1z} U_1
+
{\alpha}_{2} U_2
\right)
\nonumber\\
&&\hspace{0.2cm}
+
\frac{B^2\Phi_1}{\langle B^2 \Phi_1\rangle}
\Big(
{\alpha}_{1z}
\left\langle
B^2U_1 \Phi_1
\right\rangle
+
{\alpha}_{2}
\left\langle
B^2U_2 \Phi_1
\right\rangle
\Big)
\Bigg].
\end{eqnarray}

In order to obtain an explicit expression for $\Gamma_z$, we need the quantity $A$. So far, we have made no hypothesis about the magnetic geometry. In what follows, we assume that the stellarator has large aspect ratio; i.e. $\epsilon = a/R_0\ll 1$, where $a$ is the stellarator minor radius. Denote by $\nu_{ii*} = \nu_{ii} R_0 / v_{ti}$ the ion collisionality, where $\nu_{ii}$ is the ion-ion collision frequency and $v_{ti} = \sqrt{2T/m_i}$ is the ion thermal speed. Manipulations analogous to those employed in \cite{Helander2017b} give, for a large aspect ratio stellarator and for low collisionality ions, $\nu_{ii*} \ll \epsilon^{3/2}$,
\begin{eqnarray}\label{eq:Acdotb_final}
&&A
=
\frac{T}{Z_i e}
\frac{B}{\langle B^2 \rangle}
\Bigg[
 \left(
 {\alpha}_{1i} -\frac{3}{2} {\alpha}_{2}
\right)\Big(f_s + \langle B^2 \rangle u\Big)
\nonumber\\
&&\hspace{0.2cm}
+
\frac{f_c\Big(f_s + \left\langle B^2 u \right\rangle\Big)}{1-f_c}
({\alpha}_{1i} - 1.17 {\alpha}_{2})
\Bigg].
\end{eqnarray}
Here, $f_c$ and $f_s$ are two constants for each magnetic surface whose values can be found in \cite{Helander2017b} and the function $u$ is defined by the equation
$\bun\cdot\nabla u = 2 B^{-3}(\bun\times\nabla r)\cdot \nabla B$,
where the integration constant has been chosen so that $u(r, \theta_{\rm max}, \zeta_{\rm max}) = 0$. We have also used that the function $s$ defined in \cite{Helander2017b} is small compared to $u$ when $\epsilon \ll 1$ and we have dropped it.

Using \eq{eq:solvability_condition_3} and \eq{eq:Acdotb_final} in \eq{eq:Gamma_z_Fz1minus1_4}, we obtain the final expression for the radial impurity flux,
\begin{eqnarray}\label{eq:Gamma_z_Fz1minus1_final}
&&\Gamma_z
=
\frac{m_z {\eta} T \nu_{zi}}{Z_z e^2}
\Bigg\{
-\frac{1}{Z_z}\left\langle
B^2 \Phi_1 U_1({\alpha}_{1z}U_1 + {\alpha}_{2}U_2)
\right\rangle
\nonumber\\
&&\hspace{0.2cm}
+\frac{1}{Z_z}
\frac{\langle B^2 \Phi_1 U_1\rangle}{\langle B^2 \Phi_1\rangle}
\left\langle
B^2\Phi_1({\alpha}_{1z}U_1 + {\alpha}_{2}U_2)
\right\rangle
\nonumber\\
&&\hspace{0.2cm}
+
\frac{1}{Z_i}\Bigg\langle
U_1
\frac{B^2}{\langle B^2 \rangle}
\Bigg[
 \left(
 {\alpha}_{1i} -\frac{3}{2} {\alpha}_{2}
\right)(f_s + \langle B^2 \rangle u)
\nonumber\\
&&\hspace{0.2cm}
+
\frac{f_c(f_s + \left\langle B^2 u \right\rangle)}{1-f_c}
({\alpha}_{1i} - 1.17 {\alpha}_{2})
\Bigg]
\Bigg\rangle
\nonumber\\
&&\hspace{0.2cm}
-
\frac{1}{Z_i}
\frac{\langle B^2 \Phi_1U_1\rangle}{\langle B^2 \Phi_1\rangle}
\Bigg\langle
\frac{B^2}{\langle B^2 \rangle}
\Bigg[
 \left(
 {\alpha}_{1i} -\frac{3}{2} {\alpha}_{2}
\right)(f_s + \langle B^2 \rangle u)
\nonumber\\
&&\hspace{0.2cm}
+
\frac{f_c(f_s + \left\langle B^2 u \right\rangle)}{1-f_c}
({\alpha}_{1i} - 1.17 {\alpha}_{2})
\Bigg]
\Bigg\rangle
\Bigg\}
.
\end{eqnarray}
As advanced in Section \ref{sec:introduction}, in a stellarator the terms proportional to $E_r$ do not cancel out on the right side of \eq{eq:Gamma_z_Fz1minus1_final} in general. It can be checked that \eq{eq:Gamma_z_Fz1minus1_final} reduces to the expression for the radial impurity flux given in \cite{Helander2017} when $\varphi_1 \equiv 0$, and that in that case the right side of \eq{eq:Gamma_z_Fz1minus1_final} does not depend on $E_r$. This can be easily seen by noting that, for $\varphi_1 \equiv 0$, we have $U_1 = u$ and $U_2 \equiv 0$, and recalling that the function $s$ appearing in \cite{Helander2017} is negligible compared to $u$ when $\epsilon\ll 1$. Besides, one can prove that when \eq{eq:Gamma_z_Fz1minus1_final} is particularized for a tokamak, $E_r$ does not drive impurity flux even if $\varphi_1$ is taken into account, a result that is consistent with \cite{Helander1998}.

For $\varphi_1 \equiv 0$, the impurity flux in equation~\eq{eq:Gamma_z_Fz1minus1_final} is of order $\Gamma_z \sim (m_z n_z T \nu_{zi}/Z_z e^2 B^2) \alpha_2$, where we have used $U_2 \equiv 0$, the estimate $U_1 \sim u \sim B^{-2}$ and the fact that the terms that contain $f_s$ and $f_c$ cancel because $f_s$ and $f_c$ are flux functions. For $Z_z e|\varphi_1|/T \ll 1$, assuming that $\nabla \ln |\varphi_1| \sim a^{-1}$, we find $\Phi_1 \simeq 1 + Z_z e\varphi_1/T$ and $U_1 - u \sim U_2 \sim B^{-2} (Z_z e \varphi_1/\epsilon T)$. The scaling of $f_s$ and $f_c$ with $\epsilon$ is $f_s\sim\epsilon^{-1/2}$ and $1- f_c \sim \epsilon^{1/2}$. From \eq{eq:Gamma_z_Fz1minus1_final}, we find that $\varphi_1$ gives a correction to $\Gamma_z$ that becomes important when
\begin{equation}\label{eq:estimation_importance_varphi1}
\frac{Z_z e |\varphi_1|}{\epsilon T} \gtrsim 1. 
\end{equation}
Equation \eq{eq:estimation_importance_varphi1} means that $\varphi_1$ need not be as large as $Z_ze\varphi_1/T \sim 1$ for its effect on the impurity flux to be significant. Therefore, when $Z_z e|\varphi_1|/T \ll 1$ and \eq{eq:estimation_importance_varphi1} hold, we get the estimate
\begin{equation}
\Gamma_z \sim \left[1 + O\left(\frac{Z_z e \varphi_1}{\epsilon T}\right)\right]\frac{m_z n_z T \nu_{zi}}{Z_z e^2 B^2} \alpha_2.
\end{equation}

%%%%%%%%%%%%%%%%%%%%%%%%%%%%%%%%%%%%%%%%%%%%%%%%%
\section{Numerical evaluation of expression \eq{eq:Gamma_z_Fz1minus1_final}}
\label{sec:numerical_evaluation_Gammaz}
%%%%%%%%%%%%%%%%%%%%%%%%%%%%%%%%%%%%%%%%%%%%%%%%%

Next, we illustrate the analytical results by numerically evaluating expression \eq{eq:Gamma_z_Fz1minus1_final} in an example. We focus on an LHD configuration~\cite{Velasco2017b} at $r/a=0.8$, with major radius $R_0=3.67\,$m and minor radius $a=0.64\,$m. We take $n_i=1.2\times 10^{20}\,$m$^{-3}$, $T = 1.3\,$keV, $n'_i=-4.7\times 10^{20}\,$m$^{-4}$ and $T'=-6.9\,$keV/m. These values correspond to the plasma denoted by ``A.III" in \cite{Garcia2017}, but here we take deuterium for the main ions. In this situation, $\nu_{ii*} = 3\times 10^{-2}$. We will give numerical results for several values of $E_r$ around the ambipolar value (obtained by requiring that the radial flux of the main ions vanish), $E_r \approx -13\,$kV/m; for each case, $\varphi_1$ is calculated with the code \texttt{EUTERPE}~\cite{Kornilov2005}. For the impurities, we take $\eta' = 0$ (therefore, $\Gamma_z$ gives information on how an initially homogeneous impurity density profile tends to evolve) and give results for three  charge states of tungsten, $Z_z=24$, $Z_z=34$ and $Z_z=44$. For $Z_z = 44$, we have $\nu_{zi*} = 3.4$.

\begin{figure}
\centering
\includegraphics[width=0.7\textwidth]{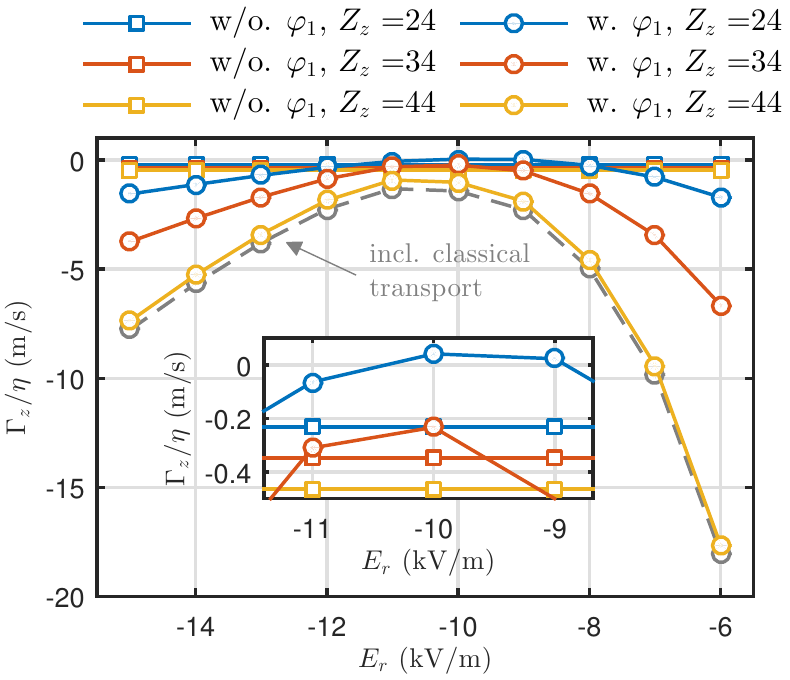}
\caption{Neoclassical impurity flux given by \eq{eq:Gamma_z_Fz1minus1_final} as a function of $E_r$ for the plasma described in the text and for $\eta' = 0$. We show pairs of curves, without $\varphi_1$ (squares) and with $\varphi_1$ (circles), for three charge states of tungsten, $Z_z = 24$, $Z_z = 34$ and $Z_z = 44$. For $Z_z = 44$, a third (dashed) curve is plotted, representing the sum of the neoclassical and classical fluxes. The inset shows a blowup of the region around $\Gamma_z/\eta = 0$ and $E_r = -10\,$kV/m.}
\label{fig:ipf_Er}
\end{figure}

In figure \ref{fig:ipf_Er} we plot the neoclassical impurity flux given by \eq{eq:Gamma_z_Fz1minus1_final} for the plasma described above, as a function of $E_r$. We show the results without and with $\varphi_1$. Without $\varphi_1$, the dependence of $\Gamma_z$ on $Z_z$ is weak and there is no dependence on $E_r$ at all. When $\varphi_1$ is included, the dependence of $\Gamma_z$ on $Z_z$ and $E_r$ is very strong. In particular, the dependence on $E_r$ is non-monotonic. In order to give an idea of the amplitude of $\varphi_1$ calculated by \texttt{EUTERPE}, we define $2\Delta\varphi_1$ as the difference between the maximum and minimum values of $\varphi_1$ on the magnetic surface. The smallest value for $\Delta\varphi_1$ in the $E_r$ scan is $e\Delta\varphi_1/T \sim 5\times 10^{-3}$, reached for $E_r \approx -12\,$kV/m, and the largest one is $e\Delta\varphi_1/T \sim 1.9\times 10^{-2}$, corresponding to $E_r \approx -6\,$kV/m. Although, in our example, impurity accumulation is almost always predicted, for $Z_z = 24$ and including $\varphi_1$ there is a region around $E_r = -10\,$kV/m where $\Gamma_z$ is less negative than in the case without $\varphi_1$, and there even exists a narrow range of values of $E_r$ where a small outward impurity flux is obtained. Thus, there exist combinations of main ion parameters, $Z_z$, $E_r$ and $\varphi_1$ that prevent impurity accumulation.

Finally, we have checked that the classical impurity flux (see an explicit expression in \cite{Buller2018}) is negligible in our example. In figure \ref{fig:ipf_Er}, the contribution of classical transport for $Z_z = 44$ is provided. The weight of classical transport is also small for the other charge states (not shown in figure \ref{fig:ipf_Er}). As argued in \cite{Buller2018}, in some stellarator configurations the ratio of classical to neoclassical transport can be larger. For instance, in configurations of Wendelstein 7-X possessing a small ratio of parallel to perpendicular electric current.

\vspace{0.2cm}

%%%%%%%%%%%%%%%%%%%%%%%%%%%%%%%%%%%%%%%%%%%%%%%%%
\section{Conclusions}
\label{sec:conclusions}
%%%%%%%%%%%%%%%%%%%%%%%%%%%%%%%%%%%%%%%%%%%%%%%%%

We have derived an explicit expression for the neoclassical radial flux of trace impurities in stellarators when the main ions have low collisionality and the impurities are collisional. This expression includes the effect of the component of the electrostatic potential that is non-constant on the magnetic surface, $\varphi_1$, which we have shown to be very strong for highly-charged impurities. In addition, in this collisionality regime, we have shown that the cases without and with $\varphi_1$ are qualitatively different: in \cite{Helander2017}, it was proven that, when $\varphi_1$ is neglected, the radial electric field does not drive impurity flux. Our calculation shows that, when the effect of $\varphi_1$ is taken into account, this does not hold any longer. We have provided realistic examples in which including $\varphi_1$ reverses the sign of the impurity flux and gives outward transport.

%%%%%%%%%%%%%%%%%%%%%%%%%%%%%%%%%%%%%%%%%%%%%%%%
\ack
%%%%%%%%%%%%%%%%%%%%%%%%%%%%%%%%%%%%%%%%%%%%%%%%

This work has been carried out within the framework of the EUROfusion Consortium and has received funding from the Euratom research and training programme 2014-2018 under grant agreement No 633053. The views and opinions expressed herein do not necessarily reflect those of the European Commission. This research was supported in part by grant ENE2015-70142-P, Ministerio de Econom\'ia y Competitividad, Spain.

%%%%%%%%%%%%%%%%%%%%%%%%%%%%%%%%%%%%%%%%%%%%%%%%
\section*{References}
%%%%%%%%%%%%%%%%%%%%%%%%%%%%%%%%%%%%%%%%%%%%%%%%

%===========================================================================

\end{document}